\documentclass[usenatbib]{mn2e}
\usepackage{psfig}

\title[An active M8.5 dwarf wide companion to LHS~4039/LHS~4040]{An active M8.5 
dwarf wide companion to the M4/DA binary LHS~4039/LHS~4040 \thanks{Based on 
observations collected with the ESO VLT, NTT and 3.6m teslescope, Chile}}
\author[R.-D. Scholz et al.]
       {R.-D. Scholz,$^{1}$\thanks{E-mail: rdscholz@aip.de} 
        N. Lodieu,$^{1}$
        R. Ibata,$^{2}$ 
        O. Bienaym\'{e},$^{2}$ 
        M. Irwin,$^{3}$ 
        M.J. McCaughrean,$^{1}$
        A. Schwope$^{1}$
        \\
        $^1$Astrophysikalisches Institut Potsdam, 
              An der Sternwarte 16, D--14482 Potsdam, Germany\\
        $^2$Observatoire de Strasbourg,
              11, rue de l'Universit\'e, F--67000 Strasbourg, France\\
        $^3$Institute of Astronomy, University of Cambridge,
              Madingley Road, Cambridge CB3 0HA, UK
            }

\date{Accepted ...
      Received ...;
      in original form ...}

\pagerange{\pageref{firstpage}--\pageref{lastpage}}
\pubyear{2003}

\begin{document}

\maketitle

\label{firstpage}

\begin{abstract}

Low-mass and brown dwarfs have  recently been found as wide companions
to many  nearby stars, formerly  believed to be single.  Wide binaries
are usually found as common  proper motion pairs. Sometimes, more than
two objects  share the same  large proper motion, identifying  them as
nearby systems. We  have found a third, low-mass  component to a known
wide binary  at a distance of  $\sim$21~pc, consisting of a  red and a
white dwarf  (LHS~4039 and LHS~4040; $\sim$150~AU  separation) The new
companion,   APMPM~J2354-3316C   separated   by   $\sim$2200~AU,   was
classified  as M8.5  dwarf.  In recent  spectroscopic observations  it
shows  a very strong  $H_{\alpha}$ emission  line and  blue continuum.
Comparing this  event to  flares in late-type  M dwarfs, we  find some
similarity with LHS~2397a, a nearby M8  dwarf which is so far the only
known example  of a low-mass star  with a tight  brown dwarf companion
(separation $<$4~AU). The level of the activity as measured by 
$L_{H_{\alpha}}/L_{bol}$ is comparable to that of the M9.5 dwarf 
2MASSW~J0149$+$29 both during the flare and in quiescence. 

\end{abstract}


\section{Introduction}

Many wide common proper motion (CPM) pairs or systems containing 3 and
more  components have  been  identified in  high  proper motion  (HPM)
surveys (e.g.  the LHS  catalogue of \citealt{luyten79}).  With proper
motions  as  large as  in  the LHS  catalogue,  i.e.   with more  than
0.5~arcsec\,yr$^{-1}$  there is  a  high probability  that  such systems  
are nearby ($d < 50$~pc) so that their angular separations of a few
arcsec to many arcmin translate to physical separations of about
$10^2..10^5$~AU.   These large separations  result in  orbital periods
from a few thousand to millions of years.

The gravitationally  bound wide  binaries and systems  are interesting
since  they  can be  considered  as  coeval.   Their large  separation
guarantees  an   independent  evolution  and   allows  an  undisturbed
investigation of the components.  CPM systems containing a white dwarf
provide the advantage that the age of the system can be estimated from
white dwarf  cooling theories.  In  addition, the question  of whether
cool  white dwarfs represent  a significant  fraction of  the Galactic
halo  dark matter  component  can  be studied  with  CPM systems  that
involving   a    white   dwarf    and   a   metal-poor    M   subdwarf
\citep{silvestri02}.

\citet{close90} estimated that  at least 3\% of local  stars should be
members of wide CPM  systems. However, they considered only relatively
bright    ($M_V<9.0$)   stars.    Recently,   many    low-mass   stars
\citep{kirkpatrick01a,lowrance02,lepine02}     and     brown    dwarfs
\citep{burgasser00,kirkpatrick01b,wilson01,gizis01,scholz03} have been
discovered as wide  companions to nearby stars of  spectral types F to
K, whereas no new stellar or sub-stellar companions have been found at
wide separations around M dwarfs within 8~pc \citep{hinz02}.

The  search for  new HPM  objects in  the southern  sky, started  as a
survey of  a few  thousand square degrees  using APM  (Automatic Plate
Measuring)  data of  UKST  (United Kingdom  Schmidt Telescope)  plates
\citep{scholz00,scholz02a,reyle02},   aimed   to   complete  the   HPM
catalogues (e.g. \citealt{luyten79}) at fainter magnitudes and thus to
find still  missing nearby low-luminosity objects.  This survey, which
was later  extended to  the whole southern  sky using  SuperCOSMOS Sky
Survey (SSS)  data \citep{hambly01a,hambly01b,hambly01c}, concentrated
on  the  search  for   nearby  free-floating  brown  dwarf  candidates
\citep{lodieu02,scholz02}.  It also led  to the  discovery of  new CPM
objects, such as one of the nearest (estimated distance 10-15~pc) cool
white  dwarf  pairs \citep{scholz02b}  and  the  nearest brown  dwarf,
$\varepsilon$~Indi~B \citep{scholz03},  with the distance  of 3.626~pc
known  from  the  Hipparcos   parallax  measurement  of  its  primary,
$\varepsilon$~Indi~A \citep{esa97}.

The  object described  here,  APMPM~J2354-3316C, was  discovered as  a
nearby field brown  dwarf candidate (based on its  large proper motion
and  photographic photometry),  and subsequently  identified as  a CPM
object to  an already known wide  binary, consisting of  a mid-M dwarf
and a  white dwarf.  To  our knowledge, this  is the first  dM/WD pair
complemented by a third, late M dwarf component with all three
components being widely separated and well suited for detailed follow-up
observations.

\section{Brown dwarf companion to red dwarf/white dwarf binary?} 

\subsection{Common proper motion}

APMPM~J2354-3316C  was first  detected  in the  HPM  survey using  APM
measurements at only two epochs in different passbands (UKST $B_J$ and
$R$ plates,  for details  of this  survey we refer  the reader  to the
paper by \citealt{scholz00}) and  later re-discovered in the SSS based
survey for nearby brown dwarf  candidates (with additional $R$ and $I$
band data  at different epochs, see  \citealt{lodieu02,scholz02} for a
brief description of  the search technique).  It can  also be found in
the recently published HPM catalogue of \citet{pokorny03}, produced by
using SSS data covering the South Galactic Cap.

The CPM pair LHS~4039/LHS~4040 was easily identified by inspecting the
finding charts at different epochs.  These stars are separated by only
about  103   arcsec  and  show   the  same  large  proper   motion  as
APMPM~J2354-3316C (about 0.5 arcsec\,yr$^{-1}$  with a position angle of about
220$^{\circ}$).  Two  more recent finding  charts, one in  the optical
$B$  band  taken   with  the  ESO  3.6m  telescope   and  one  in  the
near-infrared $K_s$ band taken from  2MASS (Two Micron All Sky Survey)
quicklook    images,   are    shown    in   Figure~\ref{fchart}    and
Figure~\ref{2massk}, respectively. The extremely red optical-to-infrared
colour of  the brown dwarf candidate  APMPM~J2354-3316C is remarkable,
especially if compared to the blue white dwarf LHS~4040.

\begin{figure}
\begin{minipage}{71mm}
\end{minipage}
\caption{Optical  finding  chart  for  APMPM~J2354-3316C  (acquisition
image  from  1999  taken  with  the  ESO 3.6m  telescope  with  a  $g$
filter). Also marked are the M4 dwarf LHS~4039 and the DA5 white dwarf
LHS~4040. }
\label{fchart}
\end{figure}

\begin{figure}
\begin{minipage}{84mm}
\end{minipage}
\caption{2MASS  $K_s$ band  quicklook image  of  APMPM~J2354-3316C. In
this near-infrared  image (epoch  1999.57), the brown  dwarf candidate
(circled) appears much brighter than its white dwarf primary, LHS~4040
(compare with Figure~\ref{fchart}).}
\label{2massk}
\end{figure}

The CPM pair L577-71/L577-72 was discovered by \citet{luyten41} during
his Bruce proper motion survey  and listed as LHS~4039/LHS~4040 in the
LHS catalogue \citep{luyten79}. \citet{luyten49} already described the
pair as  one containing a white  dwarf. A first  classification of the
blue component  as DA white  dwarf is given by  \citet{luyten52}. More
recent spectral classification was provided by \citet{oswalt88}, where
LHS~4039 and LHS~4040, are listed as dM4 and DA5+, respectively.

With additional epoch observations  now available from the 2MASS point
source catalogue \citep{cutri03} and the second release of DENIS (DEep
Near-Infrared       Survey)      data      now       available      at
http://vizier.u-strasbg.fr/viz-bin/Cat?B/denis we were able to further
improve the proper motion measurements of APMPM~J2354-3316C: Using the
multi-epoch positions given in Table~\ref{posmag} we obtained a proper
motion of
\begin{center}
$\mu_{\alpha}\cos{\delta}=-327\pm07,~~~\mu_{\delta}=-382\pm11$ [mas\,yr$^{-1}$].
\end{center}

We  also  obtained  a  new  proper  motion  measurement  for  the  two
primaries, LHS~4039  and LHS~4040, which  was complicated by  the fact
that their images were unresolved  on all Schmidt plates.  Whereas the
values  given  in \citet{reyle02}  for  the  unresolved  pair did  not
deviate  much from  Luyten's  data \citep{luyten79},  there  is a  big
difference  between the  original LHS  proper motion  and that  of the
revised LHS \citep{bakos02}.   \citet{pokorny03} have not included the
pair in their  catalogue.  Our new positions of  LHS~4039 and LHS~4040
as measured visually in the corresponding SSS FITS images with the ESO
skycat tool  are given  in Table~\ref{posmag}.  Using  these positions
plus one from  2MASS, we got the following  proper motions (confirming
the original LHS values):
\begin{center}
$\mu_{\alpha}\cos{\delta}=-307\pm08,~~~\mu_{\delta}=-410\pm08$ [mas\,yr$^{-1}$], \\
$\mu_{\alpha}\cos{\delta}=-326\pm21,~~~\mu_{\delta}=-390\pm13$ [mas\,yr$^{-1}$], \\
\end{center}
for LHS~4039 and LHS~4040, respectively.

\begin{table*}
\begin{minipage}{175mm}
 \caption[]{Positions and photometry of APMPM~J2354-3316C, LHS~4039 and
LHS~4040 at different epochs. }
\label{posmag}
 \begin{tabular}{ccccrrrr}
 \hline
Epoch & APMPM.. & LHS~4039 & LHS~4040 & APMPM.. & LHS~4039 & LHS~4040 & s \\
yr  & J2000     & J2000    & J2000    & mag     & mag    & mag  \\
\hline
1980.546 & 23 54 09.77 -33 16 19.3 & 23 54 01.56 -33 16 16.5 & 23 54 01.58  -33 16 23.2 & $B_J$=22.20 & -- & -- & 1 \\
1988.787 & 23 54 09.57 -33 16 22.2 & 23 54 01.36 -33 16 19.6 & 23 54 01.37  -33 16 26.7 &   $R$=19.24 & -- & -- & 1 \\
1994.855 & 23 54 09.39 -33 16 25.0 & 23 54 01.20 -33 16 22.3 & 23 54 01.19  -33 16 28.6 &   $I$=16.57 & -- & -- & 1 \\
1996.608 & 23 54 09.35 -33 16 25.3 & 23 54 01.18 -33 16 23.1 & 23 54 01.20  -33 16 29.5 &   $R$=19.30 & -- & -- & 1 \\
1999.574 & 23 54 09.29 -33 16 26.6 & 23 54 01.09 -33 16 24.2 & 23 54 01.07  -33 16 30.8 &   $J$=13.05 &   $J$=9.45 &   $J$=13.96 & 2 \\
         &                         &                         &                          &   $H$=12.37 &   $H$=8.91 &   $H$=13.86 & 2 \\
         &                         &                         &                          & $K_s$=11.88 & $K_s$=8.61 & $K_s$=13.73 & 2 \\
2000.682 & 23 54 09.24 -33 16 27.0 & --                  & --                   &   $I$=16.00 & -- & -- & 3 \\
         &                         &                         &                          &   $J$=13.13 & -- & -- & 3 \\
         &                         &                         &                          &   $K$=11.83 & -- & -- & 3 \\
2000.745 & 23 54 09.24 -33 16 26.7 & --                  & --                   &   $I$=16.05 & -- & -- & 3 \\
         &                         &                         &                          &   $J$=13.03 & -- & -- & 3 \\
         &                         &                         &                          &   $K$=11.96 & -- & -- & 3 \\
 \hline
 \end{tabular}
 \medskip
\end{minipage}
\begin{minipage}{175mm}
{\bf Notes: The last column gives the source of the data: 1=SSS (photographic 
magnitudes), 2=2MASS, 3=DENIS. 
Additional photometry of APMPM~J2354-3316C 
($R_C$=18.36, $I_C$=16.15, $J_s$=13.21, $H$=12.61, $K_s$=12.10)
was obtained from service observations 
with VLT-FORS1 and VLT-ISAAC in May 2000. ''--'' lacking data (merged images).}
\end{minipage}
\end{table*}

The proper motions of APMPM~J2354-3316C and LHS~4040 agree well within
the errors  whereas that of  LHS~4039 deviates from the  common proper
motion of  those two by  $\sim$20-30 mas\,yr$^{-1}$.  This is  not unexpected,
since  the orbital  motion  of  LHS~4039 around  LHS~4040  leads to  a
differential   proper  motion   of  20-24~mas\,yr$^{-1}$,   whereas   that  of
APMPM~J2354-3316C   around   LHS~4039/LHS~4040   yields   only   about
6-7~mas\,yr$^{-1}$. This assumes a circular orbits  in the plane of the sky, a
distance    of   21~pc,    0.5-0.8   solar    masses    for   LHS~4040
\citep{wegner91,bergeron95} and 0.2 and 0.1 solar masses, respectively
for LHS~4039 and APMPM~J2354-3316C.

\subsection{Photometry}

The photographic photometry  served for the preliminary classification
of  APMPM~J2354-3316C as  a brown  dwarf candidate  and  more accurate
optical   and  near-infrared   photometry  from   VLT,   2MASS,  DENIS
observations are listed in Table~\ref{posmag}. Also given is the 2MASS
photometry  for   LHS~4039/LHS~4040.  Unfortunately,  this   pair  was
unresolved in DENIS observations.

Although a  trigonometric parallax measurement of  the previously well
known pair  LHS~4039/LHS~4040 is not available,  one can independently
estimate the distance  of the individual components of  the CPM system
from   photometric  and  spectroscopic   observations.  We   obtain  a
spectroscopic  distance  of  about  15~pc for  LHS~4039  adopting  the
spectral   type   M4   \citep{oswalt88})   and  using   the   absolute
near-infrared     magnitudes     of     M4     dwarfs     given     in
\citet{kirkpatrick94}.   A  more  conservative   photometric  distance
estimate  of about 21~pc  is given  in the  ARICNS database  of nearby
stars  (http://www.ari.uni-heidelberg.de/aricns/),  where  a  spectral
type  of  M3.5  is  assigned  to  this  object.  On  the  other  hand,
\citet{silvestri01}   obviously  apply   a  distance   of   14~pc  for
LHS~4039/LHS~4040,  referring to  \citet{smith97},  but list  LHS~4039
with an  even later spectral type  of dM5, which should  place it even
closer.

From our spectroscopic classification  of APMPM~J2354-3316C as an M8.5
dwarf (see next  section) we obtain a spectroscopic  distance of about
19.5~pc  if we  compare the  2MASS  $JHK_s$ magnitudes  with the  mean
absolute magnitudes of two single M8.5 dwarfs given in \citet{dahn02}.
Our more  uncertain near-infrared classification as an  M8 dwarf would
place the object at about 25~pc distance (from comparison with four M8
dwarfs in \citealt{dahn02}).   These results for APMPM~J2354-3316C are
in   good   agreement   with   the   ARICNS   distance   estimate   of
LHS~4039.  Therefore, we  adopt the  distance  of 21~pc  for the  wide
triple system.

\section{Spectroscopic classification}

The  first optical spectra  of APMPM~J2354-3316C  were taken  with the
ESO\,3.6m/EFOSC2  camera  on  3  October  1999.   The  data  reduction
involved  subtracting  an  averaged  dark  frame and  dividing  by  an
internal flat-field  to remove fringes  above 8000\,\AA{}.  Wavelength
calibration  was made  using  He  and Ar  lines  throughout the  whole
wavelength  range.  Flux  calibration was  achieved using  an averaged
sensitivity function  created from a  spectrophotometric standard star
observed on the same night.

\begin{figure*}
\begin{minipage}{180mm}
\psfig{figure=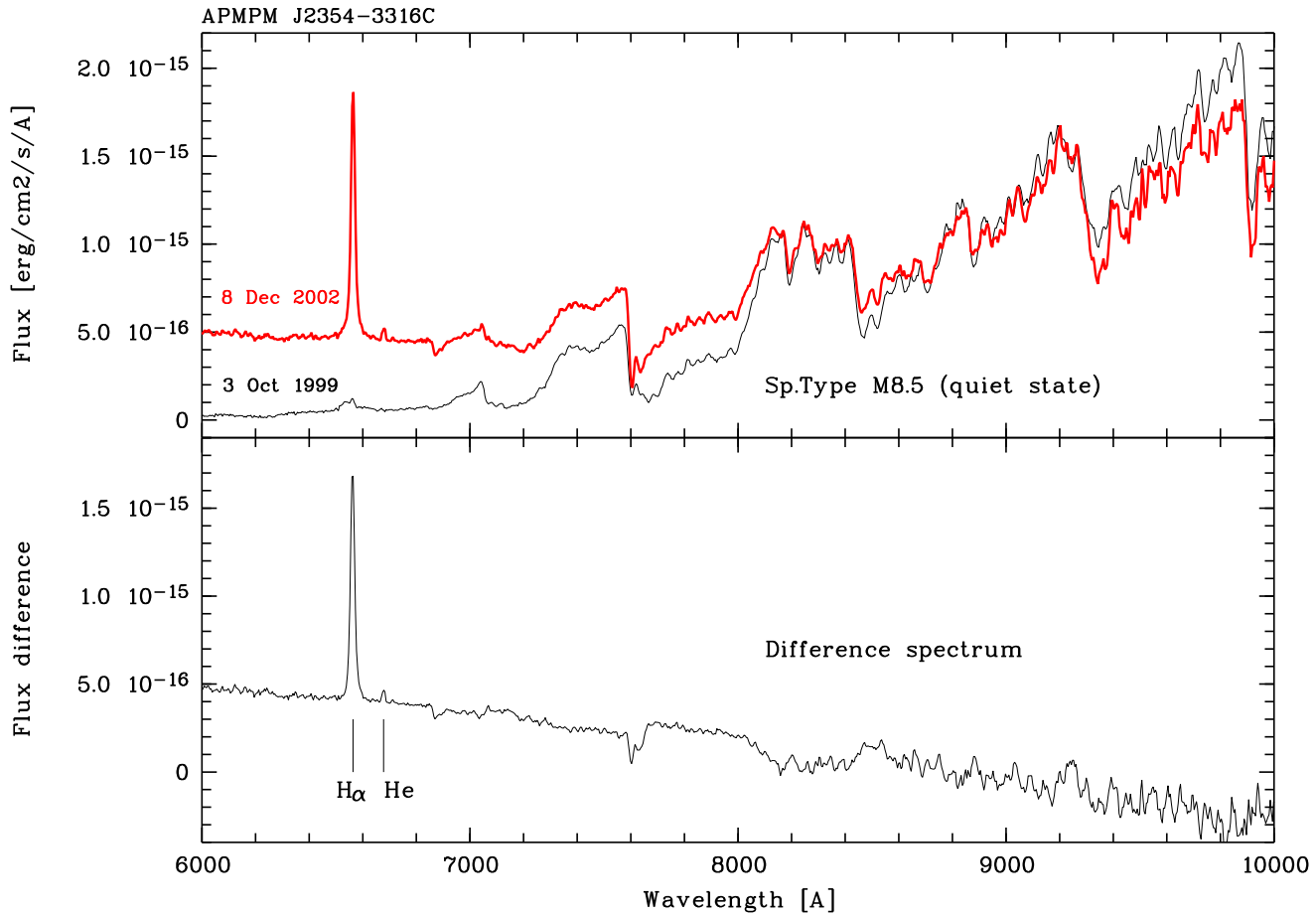,width=180mm,angle=270,clip=}
\caption{Top:  Flux calibrated spectra  of APMPM~J2354-3316C  in quiet
state (thin line)  from 1999 and with strong  $H_{\alpha}$ emission in
2002  (thick  line).  Bottom:  The  difference  spectrum,  which is  a
perfect blue veiling continuum  plus $H_{\alpha}$ at 6563.8\,\AA{} and
HeI  at   6678.1\,\AA{}  in   emission,  with  equivalent   widths  of
61.4$\pm$5.0\,\AA{} and 2.3$\pm$0.5\,\AA{}, respectively.}
\label{specdiff}
\end{minipage}
\end{figure*}

Two  optical spectra,  one  of lower  resolution  taken with  grism\#1
(4000--10000\,\AA{})    not   shown    here,   one    with   grism\#12
(6000--10000\,\AA{}) shown in  Figure~\ref{specdiff}, were similar and
typical  of  a  very late-type  M  dwarf  with  a weak  indication  of
$H_{\alpha}$  emission  in  the  grism\#12  spectrum.  In  the  latter
spectrum  (exposure time  900~s), the  object was  classified  as M8.7
using the  PC3 index  of \citet{martin99} and  as M8.3  using spectral
indices of \citet{kirkpatrick99}. Visual comparison with template spectra
(M7.5 to M9.5) from \citet{kirkpatrick99}, \citet{gizis00} and \citet{reid99}
available at http://dept.physics.upenn.edu/~inr/ultracool.html as well as
with a large number of late-type M dwarfs observed during
different runs with the same instrument setup at the ESO 3.6m
telescope (Lodieu et al., in preparation) also lead to a classification
between M8 and M9. Therefore, we  adopted a spectral type of M8.5
from the optical spectroscopy.

On 25 November 2001 we took a near-infrared spectrum (0.94-2.5~$\mu$m,
total exposure  time 300~s) of APMPM~J2354-3316C with  the SOFI camera
at the  ESO NTT.  For details  on the near-infrared  data reduction we
refer  to \citet{scholz03}.   Comparison  with near-infrared  standard
spectra          provided           by          Sandy          Leggett
(ftp://ftp.jach.hawaii.edu/pub/ukirt/skl/) led  to a classification of
M8 with  an error  of half  a subclass, which  is consistent  with the
optical classification. We prefer the optical classification since the
optical  classification  scheme  is  generally accepted  as  the  more
accurate one.

\section{A spectacular flare event}

A  new optical  spectrum  of APMPM~J2354-3316C  was  taken again  with
grism\#12  using  the  ESO\,3.6m/EFOSC2  camera  on  8  December  2002
(eposure time 540~s).  Both flux calibrated spectra are overplotted in
the  upper part of  Figure~\ref{specdiff}, whereas  the lower  part of
Figure~\ref{specdiff}  shows  the  result  after subtracting  the  old
spectrum from the new one.

The difference between  the two spectra of the  same object taken with
the  same instrument  setup  is  striking. Not  only  does the  recent
spectrum  exhibit a  very strong  $H_{\alpha}$ emission  line  with an
equivalent width  of 61.4$\pm$5.0\,\AA{}, there is also  a very strong
blue  continuum. 
Many other late-type M dwarf spectra were taken during that night,
and none of them exhibited a blue continuum, hence an instrumental effect
causing the blue continuum can be excluded. Although the instrumental
response curve is somewhat uncertain above 8500 \AA{}, we think the decline
of the difference spectrum below zero longward of 9000 \AA{} is also real. 
A similar difference spectrum was observed for 2MASSW~J0149$+$29 and the
depressed red flux attributed to an apparent increase of the opacity
during the flare \citep{liebert99}. For 2MASSW~J0149$+$29, \citet{liebert99}
could also demonstrate the weakening of the blue slope of the excess flare
continuum with time. 

Unfortunately, no  other spectrum  was taken  in the
night of  8 December 2002 so  that we do  not know how long  the event
lasted. We  assume it to  be a spectacular  flare similar to  those in
other late-type  field M  dwarfs. We do  not consider here  very young
late-type dwarfs  as observed in  clusters (e.g. \citealt{barrado02}).
An  overview of the  flare activity  at the  cool end  of the  field M
dwarfs is given by \citet{martin01}.   Among 10 known late M (M7-M9.5)
flare stars,  there seem to be  only two with a  strong blue continuum
veiling the molecular features during the flare. One of them is the M8
dwarf   LHS~2397a   \citep{bessel91},   the  other   2MASSW~J0149$+$29
\citep{liebert99}.  In contrast  to  LHS~2397a and  APMPM~J2354-3316C,
2MASSW~J0149$+$29 was  observed to have  a rich emission  spectrum and
relatively weak blue veiling. Compared to LHS~2397a, APMPM~J2354-3316C
shows  an  even   stronger  blue  veiling  and  also   a  very  strong
$H_{\alpha}$  line.  In addition  to  $H_{\alpha}$  emission, we  also
identify a HeI emission line at 6678 \AA{} with an equivalent width of
2.3$\pm$0.5\,\AA{} during the flare.

We have measured the flux of the $H_{\alpha}$ line in the flare spectrum
($2.8 \times 10^{-14}$~erg\,cm$^{-2}$\,s$^{-1}$) and in the quiescence
spectrum ($5.0 \times 10^{-16}$~erg\,cm$^{-2}$\,s$^{-1}$).
Assuming a distance of 21~pc,
we get a $L_{H_{\alpha}}$ of about $1.5 \times 10^{27}$~erg\,s$^{-1}$ during
the flare and about a factor 60 less in quiescence. Using absolute
bolometric magnitudes of two M8.5 dwarfs given in \citet{dahn02}, we get 
a mean bolometric luminosity of $L_{bol}=1.1 \times 10^{30}$~erg\,s$^{-1}$.
The resulting $L_{H_{\alpha}}/L_{bol}$ of about 0.0014 in the flare
and of about 0.000024 in quiescence are similar to the values given
by \citet{liebert99} for the M9.5 dwarf 2MASSW~J0149$+$29
(0.0025 and 0.000025, respectively).

The slope of the difference spectrum shown in Figure~\ref{specdiff} is
even steeper than in the highest flare state of 2MASSW~J0149$+$29 (see
top in Figure~3 of \citealt{liebert99}) by about a factor of 2. This
indicates an even stronger heating of the photosphere during the flare
in APMPM~J2354-3316C. The slope of the strong blue continuum in LHS~2397a
(Figure~10 in \citealt{bessel91} seems also to be steep but can not be
measured accurately due to the low signal-to-noise and the smaller
wavelength interval shown. For details on the possible origin of the blue
continuum we refer the reader to the discussion by \citet{liebert99}.         

\section{Discussion}

We have  found a  low-mass companion to  a white dwarf/red  dwarf wide
pair  at a  separation  of about  2200~AU.   The optical  spectroscopy
classifying  the  new CPM  component  as an  M8.5  dwarf  leads to  an
independent distance  estimate of 19.5~pc, which is  in good agreement
with the photometric distance estimate of the two previously known CPM
components of 21~pc given in the ARICNS database. Much closer distance
estimates of  about 15~pc (e.g.  \citealt{silvestri01,smith97}) may be
the  result of a  wrong classification  (later than  M3.5) of  the red
dwarf CPM component LHS~4039.

In  only one  of  our optical  spectroscopic  observations, the  newly
detected  object showed  a remarkably  strong $H_{\alpha}$  (plus weak
HeI) emission line together with  a very blue continuum. A similar
spectrum, although  with lower  signal-to-noise, has been  observed in
the nearby M8 dwarf, LHS~2397a, which is so far the only known example
of a low-mass star with a tight brown dwarf companion (separation less
than 4~AU) \citep{freed03}. Compared to an other
late-type (M9.5) dwarf with a spectacular flare (2MASSW~J0149$+$29),
the flare spectrum of APMPM~J2354-3316C shows an even steeper blue
continuum but a less diverse emission line spectrum.

The  spectrum obtained  at the  epoch of  the flare  event  shows some
similiarity  with the  spectrum  of a  cataclysmic  variable (CV),  it
resembles,  for instance, that  of the  low-accretion rate  dwarf nova
RBS1955 \citep{schwope02}. The complete  absence of the blue component
in APMPM~J2354-3316C at times, however, rules out a possible nature as
a  CV-like  object.  Even  for  a very  cool  white  dwarf primary,  a
hypothetical  old  CV  would  not  have  escaped  detection  with  our
low-resolution   spectroscopy.     The   accurate   (non-photographic)
multi-epoch $IJHK_s$ photometry (see Table~\ref{posmag}) does not give
a hint on strong variability.

APMPM~J2354-3316C was in the field of view of an X-ray observation with
XMM-Newton  (Revolution 369, ObsId 0103461101, PI: Aschenbach, observation
date December 13, 2001). The target of the X-ray observations was a comet
(C2000 WM1), which was detected as a bright extended X-ray and optical
source in the EPIC- and OM-images. APMPM~J2354-3316C is located behind the
bright extended structure in those images which prevents a proper source
search. However, inspection of the images reveals that no X-ray point
source is detected at the position of APMPM~J2354-3316C.
A rough upper limit of the X-ray countrate in the 0.2--12 keV band is
derived from the noise properties of the X-ray image at the target
position, $< 4\times10^{-4}$\,s$^{-1}$. Assuming an unabsorbed
Raymond-Smith plasma of 1 keV the count rate converts to
$F_X < 6\times 10^{-16}$\,erg\,cm$^{-2}$\,s$^{-1}$
and $L_X < 3 \times 10^{25}$\,erg\,s$^{-1}$. Hence, APMPM~J2354-3316C
was clearly in an inactive state at the time of the X-ray observation.

The CPM pair  LHS~4039/LHS~4040 has an estimated age  of about 1.8~Gyr
\citep{silvestri01}, and  we may assume APMPM~J2354-3316C  to have the
same age. \citet{gizis00} have  shown, that contrary to the well-known
stellar  age-activity  relationship,  older  late-type  field  stellar
dwarfs  seem to  be more  active  than younger  field (brown)  dwarfs.
However, they have also found none  of the late M dwarfs (M7 or later)
to  be  very active,  as  measured by  the  ratio  of $H_{\alpha}$  to
bolometric luminosity. In that respect, APMPM~J2354-3316C is similar 
to other late-type dwarfs, as e.g. 2MASSW~J0149$+$29.  

\section*{Acknowledgements}
We  would  like  to  thank  Hartmut  Jahrei{\ss}  for  information  on
LHS~4039/LHS~4040.  This work is  based on measurements of UKST plates
with  the APM  machine at  the Institute  of Astronomy,  University of
Cambridge,  UK and  with  the SuperCOSMOS  machine  at the  Wide-Field
Astronomy  Unit   of  the  Institute  for   Astronomy,  University  of
Edinburgh,  and on  observations  collected at  the European  Southern
Observatory,   Chile  (ESO   No.   64.H-0417,  65.L-0689,   68.C-0664,
70.C-0568).  We  acknowledge the  use of the  SIMBAD database  and the
VizieR service  operated at  the CDS Strasbourg  and of  data products
from  the  2MASS,  which is  a  joint  project  of the  University  of
Massachusetts    and   the    Infrared    Processing   and    Analysis
Center/California Institute of Technology,  funded by the NASA and the
NSF. Finally, we would like to thank the anonymous referee for his useful
comments.


\bsp 

\label{lastpage}

\end{document}